# Electromagnetic beam modulating through transformation optical structures

**Xiaofei Xu, Yijun Feng and Tian Jiang**
Department of Electronic Science and Engineering, Nanjing University,
Nanjing, 210093, China
E-mail: yjfeng@nju.edu.cn

**Abstract.** The transformation media concept based on the form-invariant Maxwell's equations under coordinate transformations has opened up new possibilities to manipulate the electromagnetic fields. In this paper we report on applying the finite-embedded coordinate transformation method to design electromagnetic beam modulating devices both in the Cartesian coordinates and in the cylindrical coordinates. By designing the material constitutive tensors of the transformation optical structures through different kinds of coordinate transformations, either the beam width of an incident Gaussian plane wave could be modulated by a slab, or the wave propagating direction of an omni-directional source could be modulated through a cylindrical shell. We present the design procedures and the full wave electromagnetic simulations that clearly confirm the performance of the proposed beam modulating devices.

## 1. Introduction

Recent progress in the research area of metamaterial has shown a great expansion of the electromagnetic (EM) wave propagation phenomenon such as negative refraction [1], perfect imaging [2], image magnification through hyper-lens [3]-[5], and the EM invisibility cloaking [6]-[16], since the electromagnetic parameters could now be arbitrary designed as desired. These extraordinary properties are directly determined by media parameters, the permittivity and the permeability. From the point of view of EM device design, it requires a certain method to regulate the material parameters to obtain the desired device properties. Recently based on the form-invariant of Maxwell's equations under certain coordinate transformations, the transformation optics proposed in [6, 8] for controlling the EM fields has been proved to be an effective approach for manipulating the material properties to satisfy the desired ray traces of the EM waves [9]. The successful application of the transformation optics to the invisibility cloaks has triggered more intensive explorations of this idea theoretically and experimentally, and many interesting theoretical results and practical approaches have been obtained [9]-[16]. Besides the invisibility cloak, EM wave concentrators, rotators and other interesting devices have also been proposed with exotic EM behaviors by utilizing the transformation optics [17]-[18]. Particularly, M. Rahm et. al., have expanded the transformation optics by the use of finite-embedded coordinate transformation which has more flexibility to the transformation design of complex materials and enabling the transfer of field manipulations from the transformation optical structures to the surrounding normal medium. Such technique has been successfully applied in the Cartesian coordinates to design a parallel beam shifter and a beam splitter [19].

In this paper, we report on applying the finite-embedded coordinate transformation method to the design of EM beam modulating devices both in the Cartesian coordinates and in the cylindrical coordinates. By designing the material constitutive tensors through different kinds of coordinate transformations, the beam width of an incident Gaussian EM wave could be modulated by a metamaterial slab, or the wave propagating direction of an omni-directional line source could be modulated through a metamaterial cylindrical shell. The finite-embedded coordinate transformation method enables these transformation optical structures to let the modulated beam unchanged while leaving the metamaterial region. The performance of the beam modulating has been verified through two dimensional (2-D) full-wave numerical simulations based on the finite element method. Such EM wave manipulating devices provide alternative ways either to expand or compress EM beam width, or to steer EM radiation to specified directions, which would find applications in optical elements, microwave antenna or other potential EM devices.

## 2. Beam Modulating in the Cartesian Coordinates

We first consider the application of the finite-embedded coordinate transformation in

Cartesian coordinates. Unlike conventional complex optical system, the beam modulating device we proposed here is composed of a slab with finite width made of transformation designed anisotropic metamaterial in which the electromagnetic waves are transformed as required. This functional material is embedded in free space with the size of only several wavelengths. To design the functional metamaterial, we utilize the so-called finite-embedded coordinate transformation method as proposed in [19]. The embedded transformations add more flexibility to the design of transformation optical structures. It enables the transfer of EM field manipulations from the transformation optical structure to the surrounding medium. The finite-embedded transform structures can also be reflectionless when certain criteria are met.

For simplicity, we restrict the problems to 2-D cases. The mathematical formalism used for the calculation of the permittivity and permeability tensors of the transformation optical structure is similar to the one reported previously in [9, 18, 19]. The transformed space filled with this structure and the original virtual space (usually is free space) are related by a possible coordinate transformation that meets the desire of EM beam modulating. To explain conveniently, we denote the transformed space as $\{x'_i\}$, and the original space as $\{x_i\}$, where $i = 1, 2, 3$ indicating the three Cartesian coordinate axes. Once the mapping from the original space to the transformed space is set, the elements of the Jacobian transformation matrix that related the two coordinate systems could be calculated as

$$J_{ij} = \frac{\partial x'_i}{\partial x_j}. \quad (1)$$

The associated relative permittivity and permeability tensors in the transformed space could be calculated as

$$\vec{\varepsilon}' = \frac{J\vec{\varepsilon}J^T}{\det(J)}, \quad (2)$$

$$\vec{\mu}' = \frac{J\vec{\mu}J^T}{\det(J)}, \quad (3)$$

where, $\vec{\varepsilon}$ and $\vec{\mu}$ tensors are those for the medium in the original virtual space and $\det(J)$ denotes the determinant of the Jacobian matrix, respectively.

Consider a slab of thickness $a$ along the $x$ axis for beam width compression and expansion. Figure 1 illustrates the mapping from the original space (here is the free space) to the transformed space with the mathematical formalism defined as the following equations,

$$\begin{cases} x' = x \\ y' = y + (\eta - 1)xy/a, \quad (0 < x' < a, \text{ within the slab}) \\ z' = z \end{cases} \quad (4)$$

$$\begin{cases} x' = x \\ y' = y, \\ z' = z \end{cases} \qquad \text{(outside the slab region)} \qquad (5)$$

where $\eta$ is the modulating coefficient denoting how much the EM beam would be modulated in the slab. From the coordinate transformation (4)-(5), the permittivity and permeability tensors of the slab could be obtained as

$$\vec{\varepsilon}' = \vec{\mu}' = \frac{1}{(1+(\eta-1)x/a)} \begin{pmatrix} 1 & (\eta-1)y/a & 0 \\ (\eta-1)y/a & ((\eta-1)y/a)^2 + (1+(\eta-1)y/a)^2 & 0 \\ 0 & 0 & 1 \end{pmatrix}. \qquad (6)$$

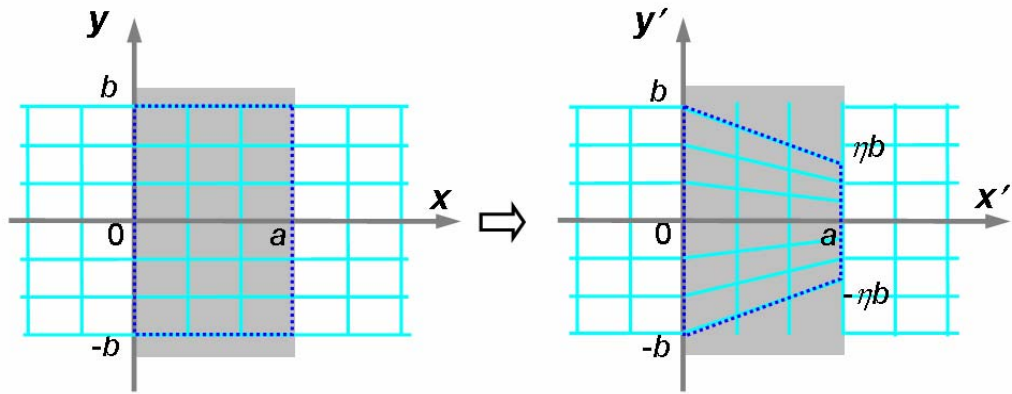

(a)

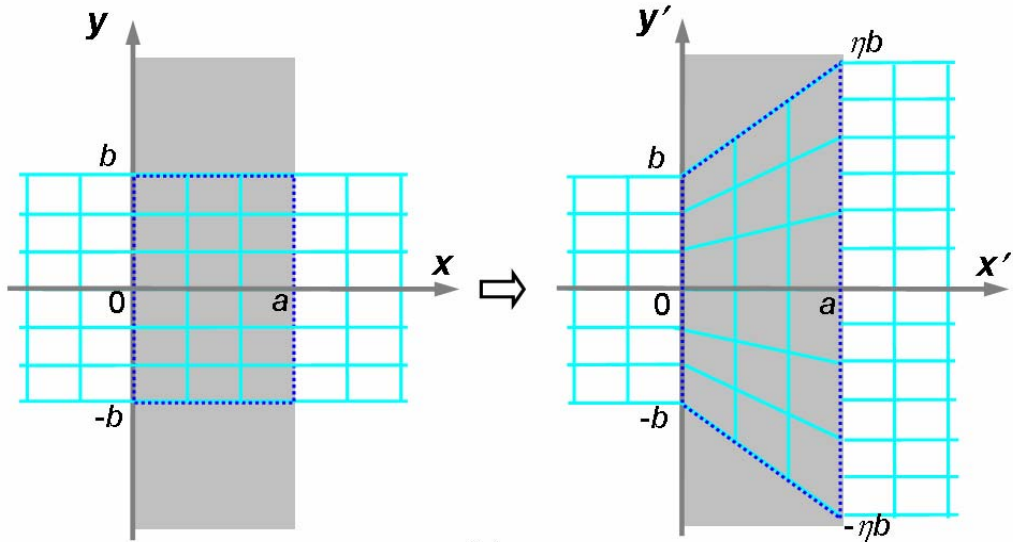

(b)

Figure 1. Coordinates mapping from the free space (left gray area) to the transformed space (a slab region with width of $a$, the right gray area) for the case of (a) traces compression or (b) traces expansion.

As indicated in figure 1, the transformation is only applied in the gray region occupied by the slab, and then the slab filled with the transformed medium is embedded in free space. Although the transformation along the right border of the slab is discontinuous as show in figure 1, as discussed in [19] the ray traces will maintain their transformed behavior when exiting the slab instead of being forced back to their original positions. This is the main difference of the embedded coordinate transformation from the conventional coordinate transformation, which excludes the abrupt change along the boundary and objectively renders the transformed medium not inherently invisible to the observer in the free space at the right side of the slab.

To confirm the performance of the proposed EM beam width modulators, we have carried out EM full wave analysis of the designed structures using a numerical simulation based on finite element method. Figure 2 illustrates the computation region for a 2-D full-wave simulation. Perfect matched layers (PML) are chosen as the surroundings for the simulation region. A transverse magnetic (TM) Gaussian plane wave (with magnetic field normal to the x-y plane) is incident either perpendicularly or obliquely into the slab from the left with a beam width $w_0$ of several wavelengths.

Since most of the energy of a Gaussian beam is restricted within $w_0$ to the beam center, we can use a slab of finite length instead of an infinite slab to modulate the Gaussian beam, which is more reasonable for practical implementation. In the simulation, the slab is chosen with width *a* along the *x* axis and length *h* along the *y* axis. To simulate the Gaussian beam modulating by a slab of finite length, *h* is chosen to be much larger than the beam width $w_0$ of the Gaussian beam.

Figure 3 shows two examples of beam width modulating by using the idea of embedded coordinate transformation. Both the transverse magnetic field and the EM power density distribution are illustrated in Figure 3. The EM wave is perpendicularly incident into the slab with a width $a = 3.33\lambda_0$ and a length $h = 26.66\lambda_0$ ($\lambda_0$ is the wavelength in free space). In figure 3 (a) and (b) the Gaussian beam with a beam width $w_0 = 5\lambda_0$ has been compressed for 50% within the transformed medium designed with a compression coefficient $\eta = 0.5$, while in figure 3 (c) and (d) the Gaussian beam with a beam width of $w_0 = 3.33\lambda_0$ has been expanded within the transformed medium designed with a expansion coefficient $\eta = 1.5$. We also notice that the simulation confirms that the EM beam remains the modulated width when exiting the transformed medium and at both sides of the slab the beam propagation are reflectionless, which is the consequence of the finite-embedded coordinate transformation method used in the design of the slab.

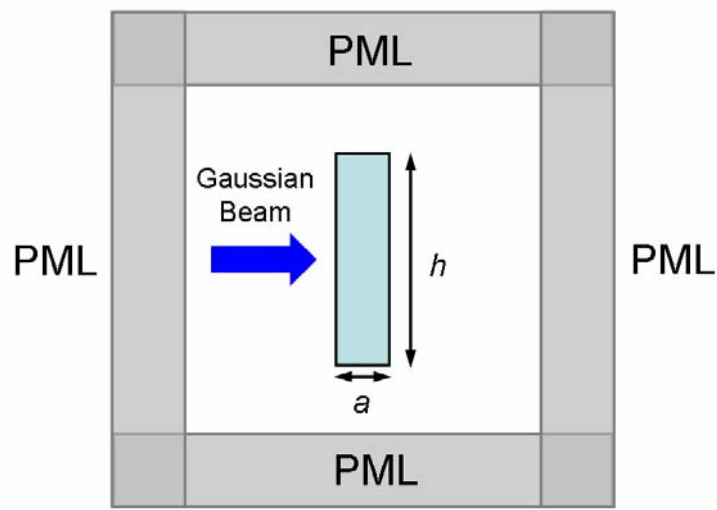

Figure 2. Computation domain and details for the full-wave simulations. A slab with finite dimension (of width *a* and length *h*, light blue colored region) is embedded in the free space which is surrounded by the PML (gray regions). A Gaussian plane wave is considered to impinge to the slab.

The transformation optical structures should also be effective for any oblique incident waves, since the design procedure does not depend on the incident angles. This has been confirmed in figure 4 which shows the beam modulating for an oblique incident Gaussian plane wave. The physical sizes of the slab are the same as that discussed in figure 3. Either a beam compression ($\eta = 0.75$) or a beam expansion ($\eta = 1.25$) has been clearly demonstrated in figure 4 (a) and (b) or (c) and (d), respectively.

The reflectionless characteristics of the proposed beam modulators also enable us to cascade them to obtain large compression or expansion of the EM power density. Figure 5 shows an example of beam width compression through two cascaded slabs resulting in more than 4 times enhancement of the output EM power density, which could be applicable to some optical elements.

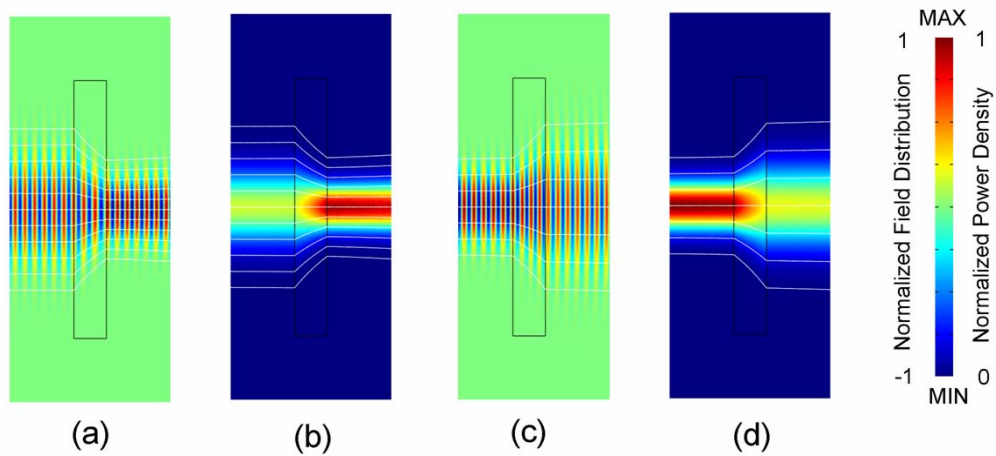

Figure 3. Beam modulating of a perpendicular incidence wave by a slab. The normalized transverse magnetic field distribution, power flow lines (white lines) [(a) and (c)] and normalized power density [(b) and (d)] for a beam width compression [(a) and (b)], or a beam width expansion [(c) and (d)].

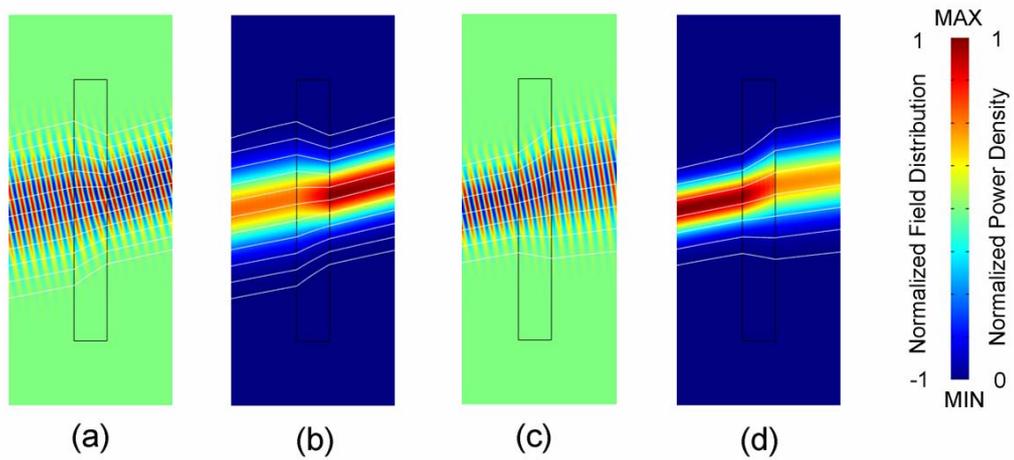

Figure 4. Beam modulating of an oblique incidence wave by a slab. The normalized transverse magnetic field distribution, power flow lines (white lines) [(a) and (c)] and normalized power density [(b) and (d)] for a beam width compression [(a) and (b)], or a beam width expansion [(c) and (d)].

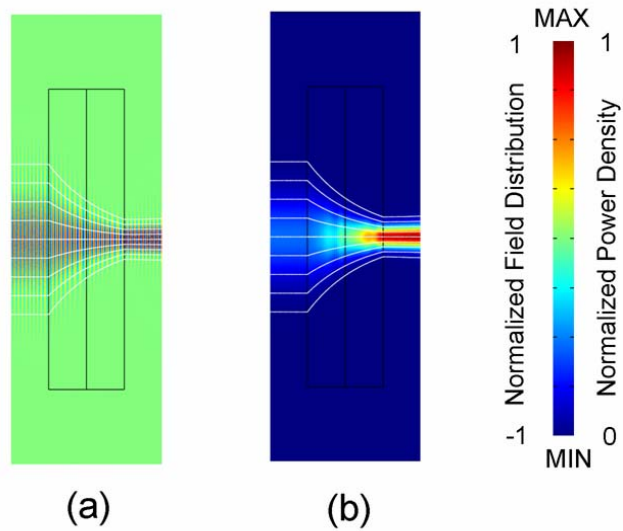

Figure 5. The normalized transverse magnetic field distribution (a), power flow lines (white lines) and normalized power density (b) for a beam width compression by two cascaded slabs.

## 3. Beam Modulating in the Cylindrical Coordinates

Next we extend the above idea of beam modulating to the cylindrical coordinates. The finite-embedded coordinate transformation method is now applied to the cylindrical system. Instead of EM traces modulating in the *y* direction as discussed in the previous section, we consider beam modulation in the circumferential direction in the cylindrical system. By filling a shell region with transformation designed functional material that surrounding an EM source, such beam modulation in the circumferential direction could be used to squeeze an omni-directional cylindrical wave to some special directions, either to increase the radiation directivity of the source or to render the original source confused to the outsider observer.

To modulate the cylindrical wave, a coordinate transformation is required between two cylindrical coordinates, under which the uniformly distributed radial traces in the original cylindrical system could be squeezed in the circumferential direction to redistribute within certain specified directions in the transformed cylindrical system. Different mappings between the two coordinates will yield different beam modulating effects and result in different field distributions. Figure 6 describes one of the 2-D mappings from a cylindrical shell region (gray region in the left) in the original virtual space to a shell region (gray region in the right) in the coordinate transformed space. Generally, the coordinate transformation formalisms could be described as

$$\begin{cases} \rho' = \rho \\ \theta' = F(\rho, \theta, a, b), \quad (a < \rho < b \text{, within the shell region}) \\ z' = z \end{cases} \quad (7)$$

$$\begin{cases} \rho' = \rho \\ \theta' = \theta, \quad \quad \quad \quad \quad \text{(outside the shell region)} \\ z' = z \end{cases} \quad (8)$$

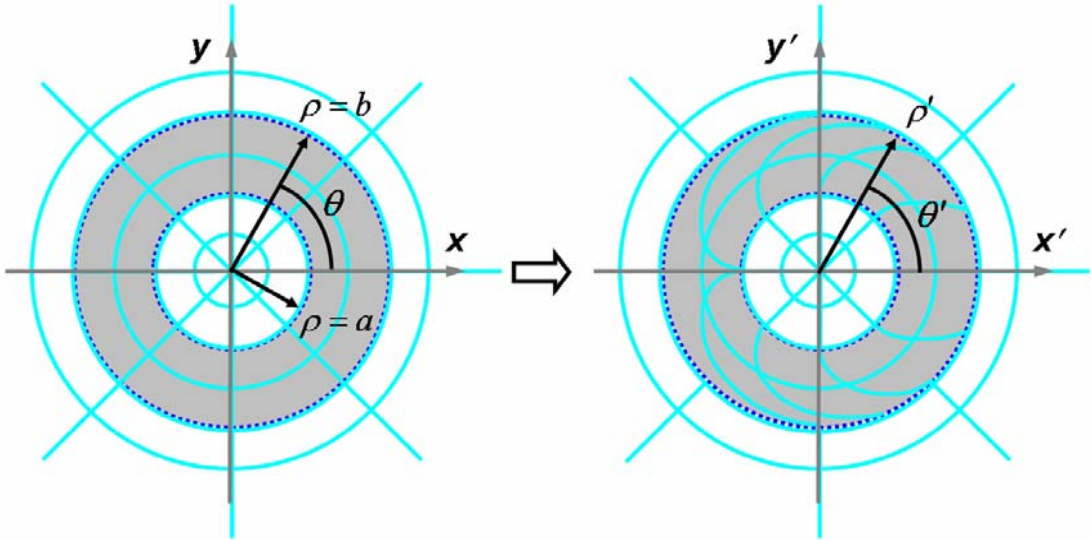

Figure 6. Coordinate mapping from a cylindrical shell region (gray region in the left) in the original virtual space to a shell region (gray region in the right) in the coordinate transformed space with the outer radius $b$ and inner radius $a$. the uniformly distributed radial traces within $\pi \leq \theta \leq -\pi$ in the original shell have been squeezed into $\pi/2 \leq \theta' \leq -\pi/2$ within the shell.

Where, $F(\rho,\theta,a,b)$ is a specified function that determines the transformation of the space in the circumferential direction and it should satisfy $F(\rho,\theta,a,b)|_{\rho=a} = \theta$ to ensure the continuity of the radial traces at the inner boundary of the shell. However, no limitation is required for the angle coordinate at the outer boundary. Therefore the medium property for the transformed region could be designed by applying the finite-embedded coordinate transformation method to this region and then embedded in the free space similar to the previous section.

To demonstrate the design procedure, one simple example of the transformation is given in the following,

$$\begin{cases} \rho' = \rho \\ \theta' = a\theta/\rho, \\ z' = z \end{cases} \quad \text{(within the shell region)} \tag{9}$$

$$\begin{cases} \rho' = \rho \\ \theta' = \theta . \\ z' = z \end{cases} \quad \text{(outside the shell region)} \tag{10}$$

If $b = 2a$, the above transformation indicates that the uniformly distributed radial traces within $\pi \le \theta \le -\pi$ in the original shell have been squeezed into $\pi/2 \le \theta' \le -\pi/2$ upon the outer boundary of the transformed shell as shown in figure 6. The material constitutive parameters of the shell in the transformed space are derived through the Jacobian transformation matrix

$$\boldsymbol{J}(\rho,\theta,z) = \begin{pmatrix} 1 & 0 & 0 \\ -a\theta/\rho & a/\rho & 0 \\ 0 & 0 & 1 \end{pmatrix}, \tag{11}$$

and the permittivity tensor is obtained through (2)-(3) as

$$\vec{\varepsilon}' = \begin{pmatrix} \varepsilon'_{\rho\rho} & \varepsilon'_{\rho\theta} & \varepsilon'_{\rho z} \\ \varepsilon'_{\theta\rho} & \varepsilon'_{\theta\theta} & \varepsilon'_{\theta z} \\ \varepsilon'_{z\rho} & \varepsilon'_{z\theta} & \varepsilon'_{zz} \end{pmatrix} = (\rho/a) \begin{pmatrix} 1 & -a\theta/\rho & 0 \\ -a\theta/\rho & (a\theta/\rho)^2 + (a/\rho)^2 & 0 \\ 0 & 0 & 1 \end{pmatrix}. \tag{12}$$

Representing in the Cartesian coordinates, they become

$$\vec{\varepsilon}'(x',y',z') = \begin{pmatrix} \varepsilon'_{xx} & \varepsilon'_{xy} & \varepsilon'_{xz} \\ \varepsilon'_{yx} & \varepsilon'_{yy} & \varepsilon'_{yz} \\ \varepsilon'_{zx} & \varepsilon'_{zy} & \varepsilon'_{zz} \end{pmatrix}, \tag{13}$$

where, the tensor elements are

$$\begin{aligned}
\varepsilon'_{xx} &= \varepsilon'_{\rho\rho}\cos(\theta')^2 + \varepsilon'_{\theta\theta}\sin(\theta')^2 - \varepsilon'_{\rho\theta}\sin(\theta')\cos(\theta') - \varepsilon'_{\theta\rho}\sin(\theta')\cos(\theta') \\
\varepsilon'_{yy} &= \varepsilon'_{\rho\rho}\sin(\theta')^2 + \varepsilon'_{\theta\theta}\cos(\theta')^2 + \varepsilon'_{\rho\theta}\sin(\theta')\cos(\theta') + \varepsilon'_{\theta\rho}\sin(\theta')\cos(\theta') \\
\varepsilon'_{xy} &= (\varepsilon'_{\rho\rho}-\varepsilon'_{\theta\theta})\sin(\theta')\cos(\theta') + \varepsilon'_{\rho\theta}\cos(\theta')^2 - \varepsilon'_{\theta\rho}\sin(\theta')^2 \\
\varepsilon'_{yx} &= (\varepsilon'_{\rho\rho}-\varepsilon'_{\theta\theta})\sin(\theta')\cos(\theta') + \varepsilon'_{\theta\rho}\cos(\theta')^2 - \varepsilon'_{\rho\theta}\sin(\theta')^2 \\
\varepsilon'_{zx} &= \varepsilon'_{zy} = \varepsilon'_{xz} = \varepsilon'_{yz} = 0 \\
\varepsilon'_{zz} &= \rho'/a
\end{aligned} \quad (14)$$

The permeability tensor components are of the same form as those of the permittivity. Similar discontinuity happens at the outer boundary of the shell due to the use the finite-embedded coordinate transformation method.

To verify the performance of the beam modulating in the circumferential direction, full-wave simulations have been carried out and displayed in figure 7. A linear line source is located in the origin which radiates cylindrical waves omni-directionally in the free space (figure 7 (a) and (b)). To modulate the beam, the line source is surrounded by a shell with the outer radius $b = 3.33\lambda_0$ and inner radius $a = b/2 = 1.67\lambda_0$, which is filled with the transformed medium designed through (12)-(14). As shown in figure 7 (c) and (d) the omni-directional cylindrical waves radiated from the line source are condensed into the right half space leaving the other half space with almost no radiation due to the transformed medium in the shell. The squeezed beam keeps the bended behavior when exiting the shell into the outer free space. Due to the non-uniform coordinate transformation formalism of (9), the EM power density is not distributed uniformly in the shell with more power squeezed near the directions of $\theta' = \pm\pi/2$. The simulation result clearly demonstrates the beam modulation from omni-directional to radiation in the half space as desired from the transformation (9) at $a = b/2$.

To demonstrate the arbitrary beam modulation ability by using the transformation designed shell, three examples have been given in figure 8 showing different modulating effects. The omni-directional radiation could be split into double beams (figure 8 (a) and (b)), treble beams (figure 8 (c) and (d)) or even quadruple beams (figure 8 (e) and (f)) radiating to four perpendicular directions. Both the near-field distribution and the power density have clearly confirmed the different designs of modulating beams from omni-directionality to some special directionality. The transformation optical structures is designed based on the geometrical optics and the ray tracing, therefore small discrepancy exists between the design and the simulation due to some undesired dispersion of the EM fields resulted from the diffractions.

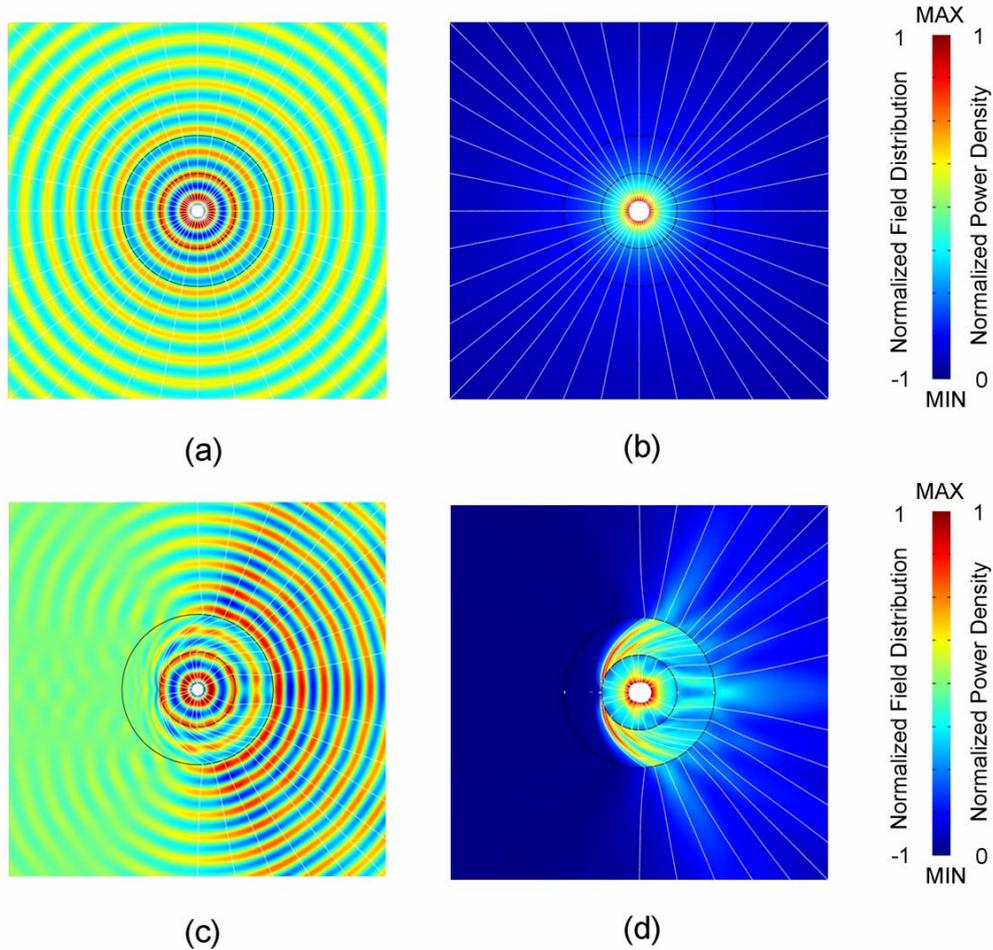

Figure 7. Transverse electric field distribution, power flow lines (white lines) [(a) (c)] and power density [(b) (d)] of a line source at the origin radiating in the free space [(a), (b)] and in the transformed space with a transformation designed shell surrounding it [(c), (d)].

4. **Conclusions**

In this paper, we have successfully applied the finite-embedded coordinate transformation method to design EM beam modulating devices both in the Cartesian coordinates and in the cylindrical coordinates. Transformation optical structures have been designed by determining the permittivity and permeability tensors of the structures through different kinds of coordinate transformations. In the Cartesian coordinates, the beam width of the Gaussian plane waves could be either compressed or expanded by a slab of the transformation optical structure, while in the cylindrical coordinates, the radiation of an omni-directional line source could be modulated to specified directions through a cylindrical shell of the transformation optical structures. Both the EM field distributions and the power density flows through 2-D full-wave numerical simulations have confirmed the performance of the beam modulating. Such EM wave manipulating devices and the design method could not only provide alternative ways either to expand or compress EM beam width, or to squeeze EM radiation to particular directions, but also find applications in optical elements, microwave antenna system or other potential EM devices.

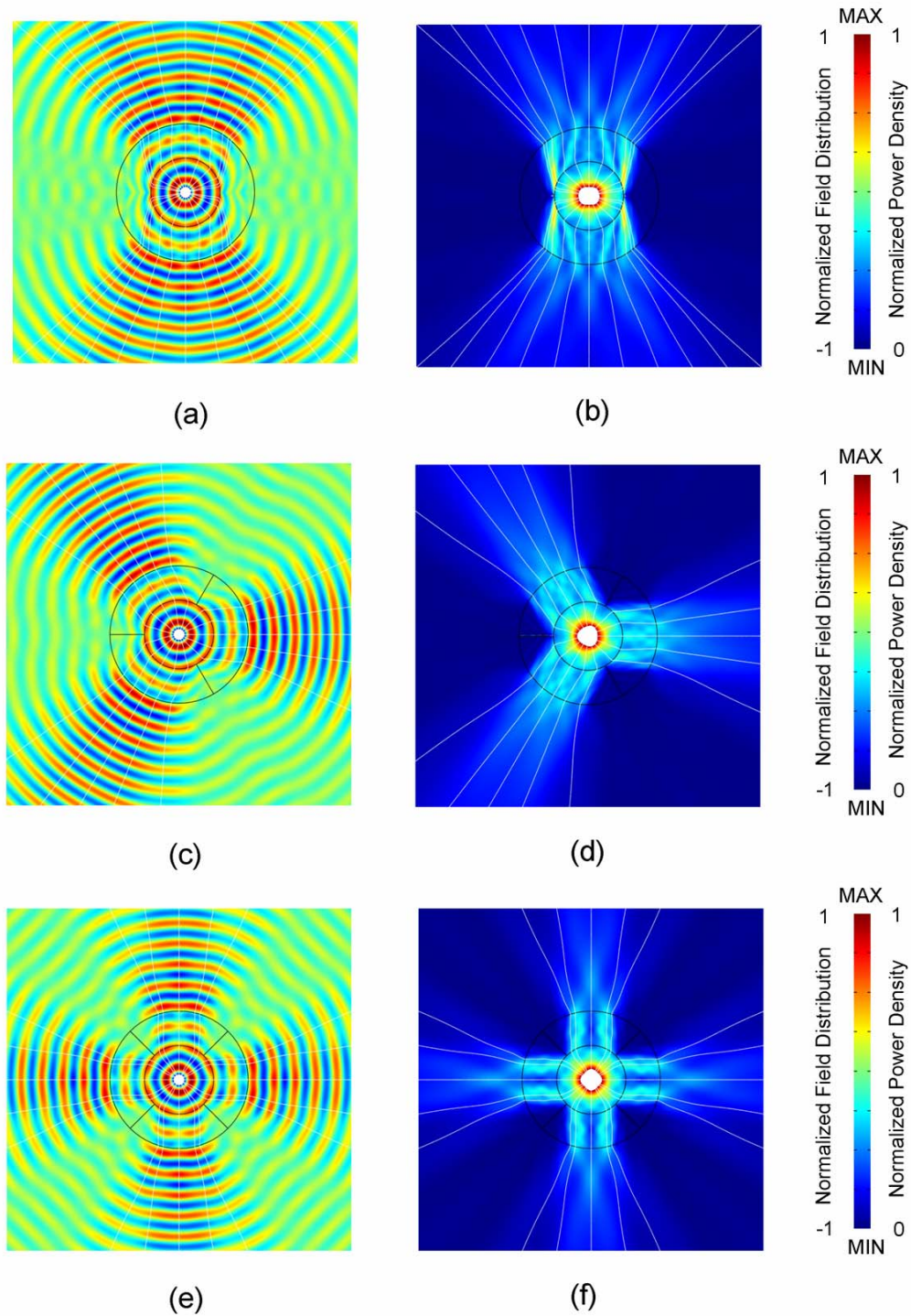

Figure 8. Transverse electric field distribution, power flow lines (white lines) [(a), (c), (e)] and the power density distribution [(b), (d), (e)] for a double beam splitter [(a), (b)], a treble beam splitter [(c), (d)] and a quadruple beam splitter [(e), (f)] (the black lines denote the boundary of the shell filled with the transformation optical structures).

**Acknowledgements**

This work is supported by the National Basic Research Program of China (2004CB719800), and the National Nature Science Foundation of China (No. 60671002).

reflectionless complex media by finite embedded coordinate transformations *Phys. Rev. Lett.* **100** 063903